\documentstyle[11pt,paspconf,epsf]{article}
\index{Circumnuclear disk}
\index{mini-spiral}
\index{Northern arm}
\index{gas dynamics}
\index{star formation near black hole}
\index{orbits in Galactic Center potential}
\begin{document}
\title{The circumnuclear disk and ionized gas filaments as remnants of
tidally disrupted clouds}
\author{R.H. Sanders}
\affil{Kapteyn Astronomical Institute, Groningen, The Netherlands}
\begin{abstract}
Sticky particle calculations indicate that a coherent structure, a dispersion
ring, forms when a cloud on a low angular
momentum orbit passes close to the dynamical center of a potential containing
a point mass.  The cloud is tidally stretched and differentially wrapped,
and dissipation in shocks organizes the gas into a precessing
off-set elliptical ring which can persist for many rotation periods.  
The morphology and kinematics of the 
circumnuclear disk (CND) between 2 and 5 pc and the Northern arm in the 
inner 1 pc are well-represented by such structures.  In the case of
the Northern Arm, strong shocks which arise during the formation of
the dispersion ring can lead to star formation even in the near tidal
field of a massive black hole.    

\end{abstract}

\section{The relation between gas flow and periodic orbits}

To what extent can the gas structures in the inner few parsecs of the
Galaxy be understood in terms of gas flow in a gravitational field?
I am referring to the well-studied circumnuclear disk (CND) and the
ionized gas filaments within the central cavity of the CND (Morris
\& Serabyn 1996, Mezger et al. 1996).  The morphology and kinematics
of the ionized gas filaments have been modeled by motion along 
Keplerian orbits appropriately projected on the plane of the sky
(Serabyn et al. 1988, Herbst et al. 1993, Roberts et al. 1996); 
the success of such models strongly suggests that the motion is primarily
orbital and that other possible mechanisms (e.g., 
stellar winds, magnetic fields, supernova explosions) do not significantly
influence the overall, systematic pattern of flow.  But the material is,
afterall, gaseous, and how well can gas motion be approximated by
orbital motion?

It is obvious that the hydrodynamic equation of 
motion written in Lagrangian form 
is the equation of motion of a particle when the pressure gradient and
viscous stress terms are
omitted.  This means that, in a gravitational field, if pressure
forces are negligible (including thermal, turbulent, viscous, and 
magnetic), then gas elements move on orbits.  Moreover, for steady
state inviscid gas flow in a gravitational field the gas streamlines 
are non-intersecting periodic orbits.  Of course, not all orbits can be
streamlines because orbits may loop and there cannot be two values of
the fluid velocity at one point.  

Fig.\ 1a shows a typical orbit in a gravitational field resulting from
a point mass embedded in an approximate isothermal sphere.  This 
mass distribution is chosen to mimic that implied by stellar kinematics
in the inner few parsecs of the Galaxy,
as described by Eckart et al. and Ghez et al. in this volume (the details
are given by Sanders 1998, Paper 1).  The orbit is typical of that in a 
general axisymmetric potential, but the hard core, the point mass, causes an 
abrupt bending of the orbit at closest approach (strong scattering).
Now one might say that such an orbit could
not possibly be a gas stream line because of the loops.  However, it is
possible to find a rotating frame in which the orbit appears as in 
Fig.\ 1b; this is a possible gas streamline.  It was the proposal of
Lindblad (1956, in connection with the problem of spiral structure) 
that a gas cloud on such an orbit would disperse to
form such a structure which would precess at a fixed angular velocity,
hence the term ``dispersion ring''.
In this case, the precession would be counter to the sense of particle
motion with an angular velocity of -26 km s$^{-1}$pc$^{-1}$.

\begin{figure}
\plottwo{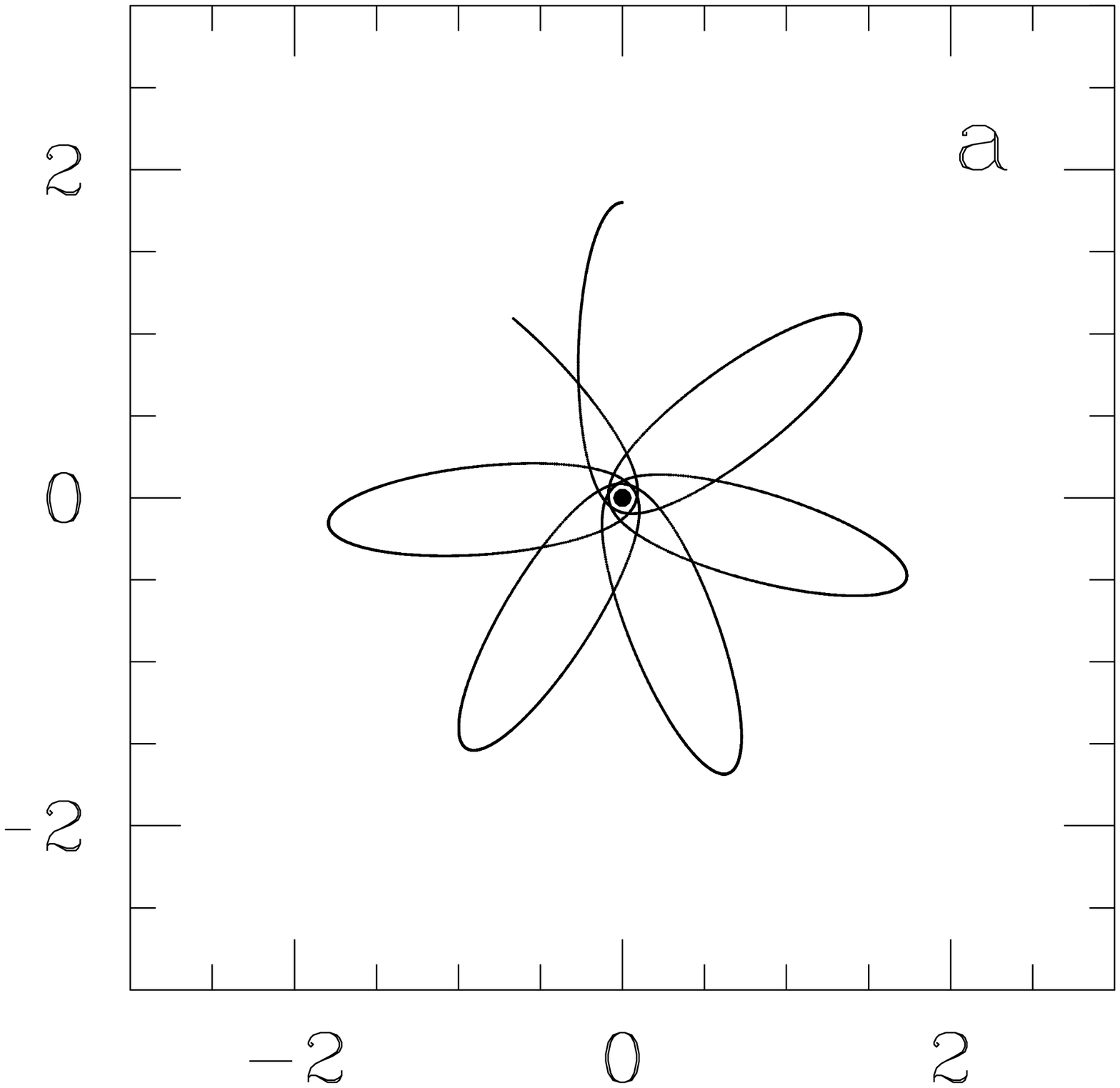}{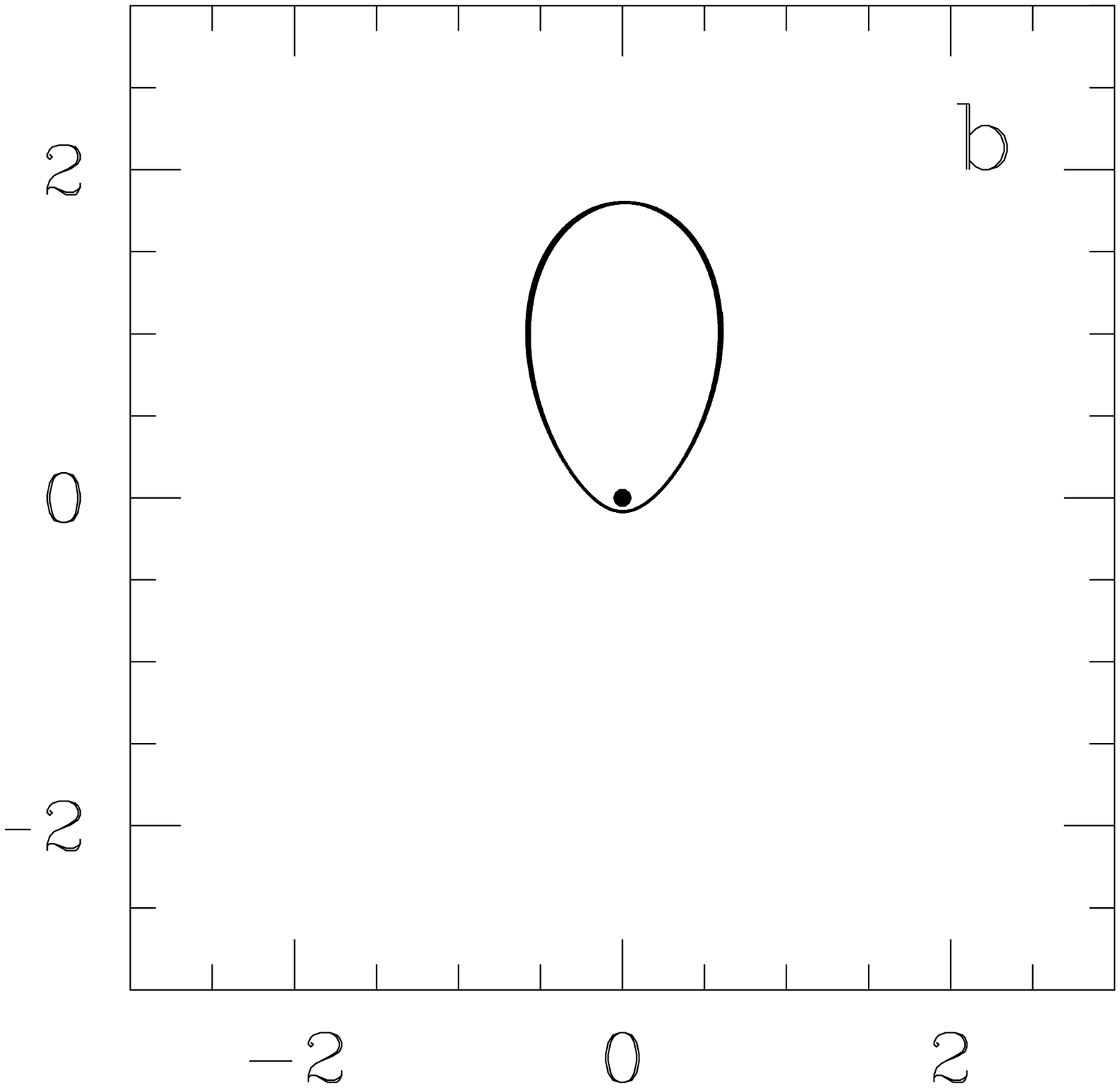}
\plottwo{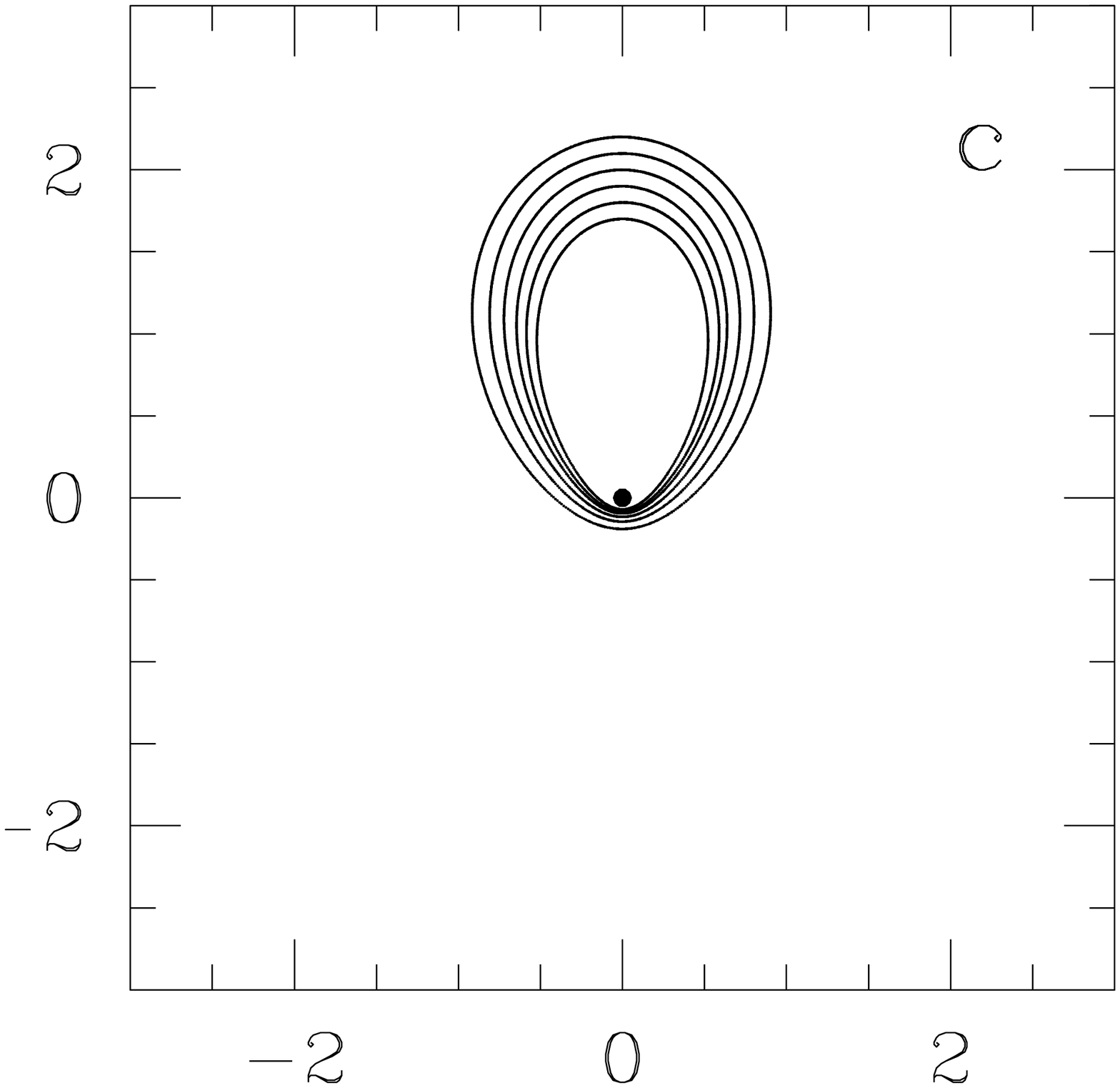}{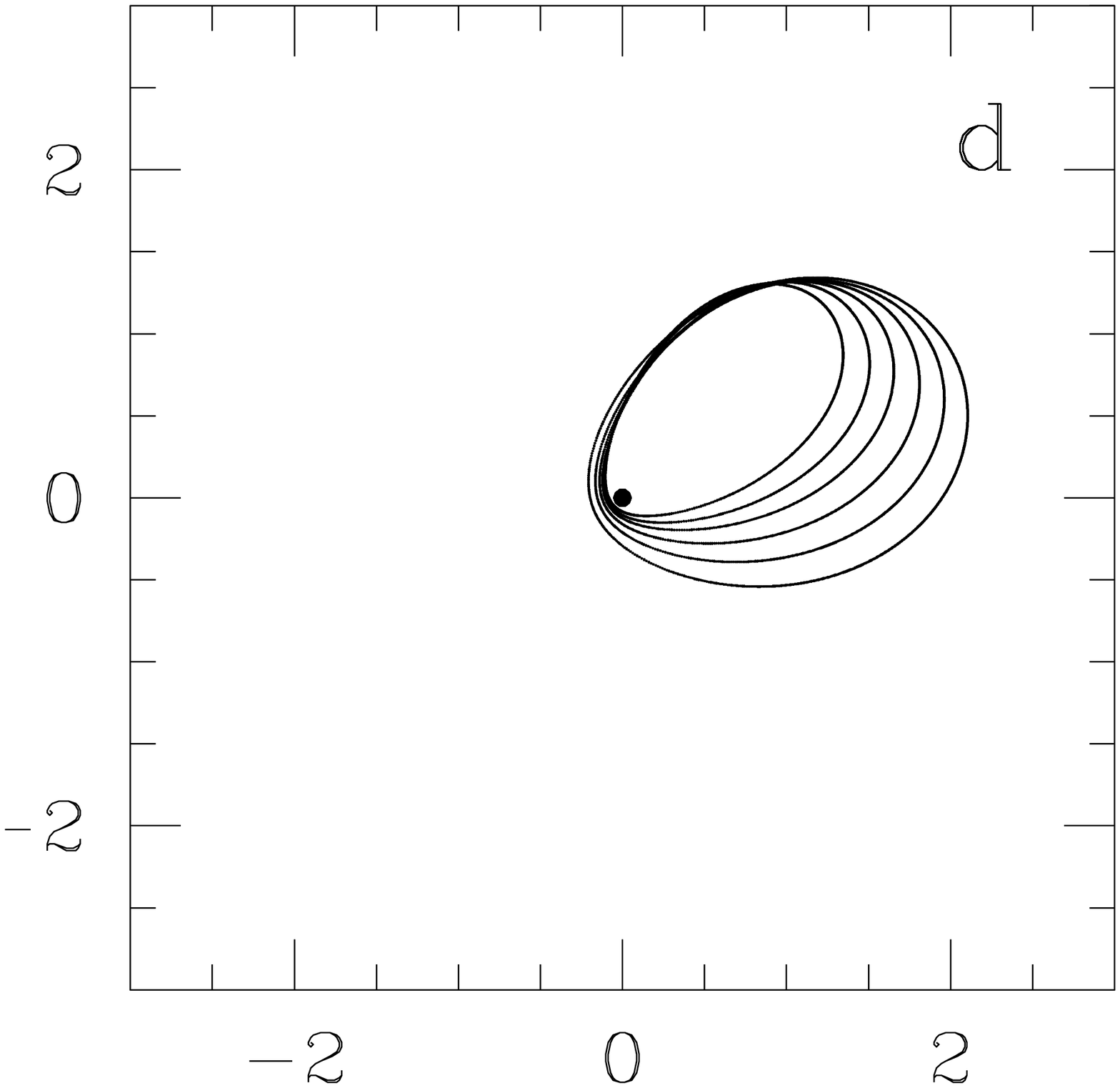}
\caption{a. An orbit in the gravitational field of an isothermal
sphere containing a point mass.  The initial distance from the center
is 1.8 pc and the tangential velocity is 20\% of that required for a 
circular orbit.  Distance is in parsecs.   b.  The same orbit in a 
frame rotating counter to the sense of particle motion with $\Omega_p
= -26$ km s$^{-1}$pc$^{-1}$.  c.  A collection of such orbits in the 
rotating frame
which could serve as gas streamlines for a persisting oval structure
in the Galactic Center.  d.  The same structure after several orbital
periods; the appearance has changed due to differential precession.
In all cases here the sense of particle motion is counter-clockwise and
the precession is clockwise.}

\end{figure}

One could find a set of such orbits covering a range of energy,
all of which precess at about the same rate.  Such a configuration is
shown in Fig.\ 1c.  For these orbits
the maximum distance to the center varies between
1.7 and 2.2 pc and the tangential velocity at this point varies between
0.18 and 0.27 times the circular velocity.  It is evident that these
``streamlines'' crowd at the point of closest approach to the point mass 
and this would cause a density enhancement there (in spite of the
higher velocities).  It is actually impossible to maintain this structure
forever because the outermost orbits precess (or rather, regress) a bit
faster than the inner most orbits (remember, nearer the center the 
potential is closer to Keplerian).  So after several orbital periods
we might expect the configuration to appear as it does in Fig.\ 1d.
Here we see that the orbits have crowded on the leading edge which 
would give the appearance of a one-arm spiral.  In fact, the orbits
have begun to cross which means that dissipative forces (shocks) will
intervene and the orbits can no longer represent streamlines.
This will cause a loss of energy and the gas will settle, on some timescale,
to a circular ring at a radius appropriate to the specific 
angular momentum.  

\section{The fate of clouds on low angular momentum orbits}

Of course I can create a structure like this artificially by loading
gas on such streamlines, and it would
persist for much longer than a characteristic orbital period.  But
how might such a structure arise naturally?  In particular, how might
it arise in the inner few parsecs of the Galaxy?  Let us consider the
fate of a gas cloud launched on a low angular momentum orbit-- an 
orbit which will carry the cloud near the central point mass.  We will
look at two cases:  1)  A clumpy cloud initially between five and ten 
parsecs from the center on an orbit which will carry it to within one or two
parsec of the center (initial tangential velocity, $V_t$, is 
0.4 of the circular 
velocity, $V_c$).  2)  A small cloud in a very low angular momentum orbit;  
i.e., initially between two and three parsecs on an orbit which takes it to
within 0.1 to 0.2 pc of the center ($V_t = 0.2 V_c$).  The first case
may be relevant to the circumnuclear disk (CND) and the second to
the Northern Arm.

I have done these calculations by means of a sticky particle code (Paper 1) 
in which the motion of 4000 particles is computed in the gravitational 
potential described above.  The effects of dissipation are included by
allowing particles to interact:  two particles exert a force on one
another proportional to their velocity difference, but only if they
are approaching.  The procedure is similar to SPH codes but ordinary gas
pressure forces are neglected; this mimics a bulk viscosity which is only 
effective if the flow is converging.  Similar schemes have been used in
simulations of accretion disks (e.g., Syer \& Clark 1992);  
the rationale is that dissipation in
strong shocks is the only important gas dynamical effect if the flow
is highly supersonic (as it is).  

In the first case (the ``CND'' cloud), the time evolution 
is described in Paper 1.
After several rotation periods, the material from the disrupted cloud has
formed an offset asymmetric ring with an extended arm.  The entire structure
is precessing at a mean rate of about -12 km s$^{-1}$pc$^{-1}$ 
(counter to the sense
of rotation).  Gradually, on a timescale of two to three million years
the structure circularizes and the cavity shrinks.  The reason for
this circularization over a timescale of about 10 rotation periods, is 
differential precession.  Over the rather large range in radius covered
by the cloud, the outer parts precess more slowly than the inner parts
leading to crossing of streamlines and strong dissipation in shocks.
Non-the-less, the structure, as it appears 0.85 to 1 million years
after the initial passage, 
does bear a strong resemblance to the CND, both in morphology
and kinematics.  This is shown in Fig.\ 2 as a particle plot
with proper motion vectors projected onto the plane of the sky.
Note the asymmetric central cavity; because the streamlines 
crowd on the side 
nearest the point mass, the highest gas density is observed there
(north-western side).  On the opposite side the ring is quite tenuous
just as observed in the actual CND (G\"usten et al. 1987)
\begin{figure}
\plotone{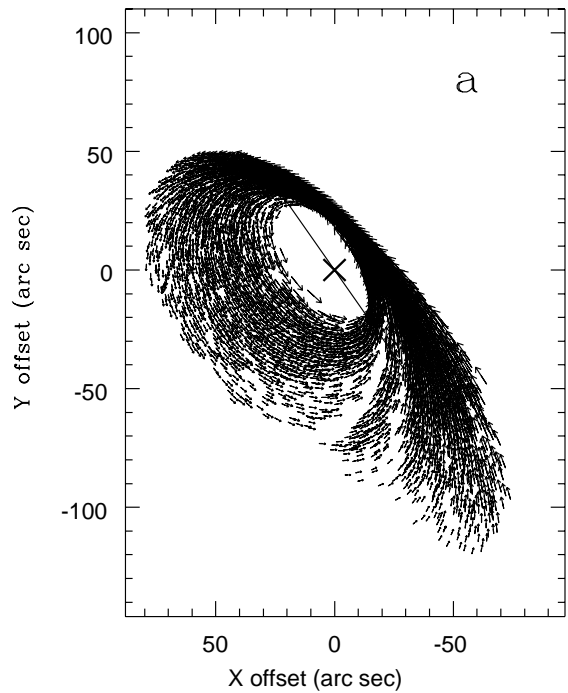}
\caption{The projected CND model (from Paper 1) $8.5\times 10^5$ years 
(three orbital periods) after launching the
clumpy cloud on its low angular momentum orbit.  
The orbital plane is 
inclined 60 degrees with respect to the plane of the sky and the
position angle of the line-of-nodes (solid line) is 35 degrees.
The north-western side (upper right) is the near side in this projection.  
Units are in arc seconds offset from the dynamical center, presumably
Sgr A$^*$.  This is a particle plot where the 
the arrows at the location of the particles indicate the sense 
and magnitude of proper motion.  The asymmetric central cavity and
the higher gas density on the side nearest the dynamical center, 
observed characteristics of the CND, are
model-independent properties of a dispersion ring.}
\end{figure}

The dispersion ring model is not entirely relevant to this model for the
CND;  there is just too much differential precession over the width
of the ring.  However, the structure does persist considerably longer
than a characteristic rotation period ($2.5\times 10^5$ years at 
r = 3.5 pc).  The important point is that the tidal stretching and
differential wrapping of an initially clumpy cloud is consistent with the
observed kinematics and morphology of the CND.

\begin{figure}
\plotone{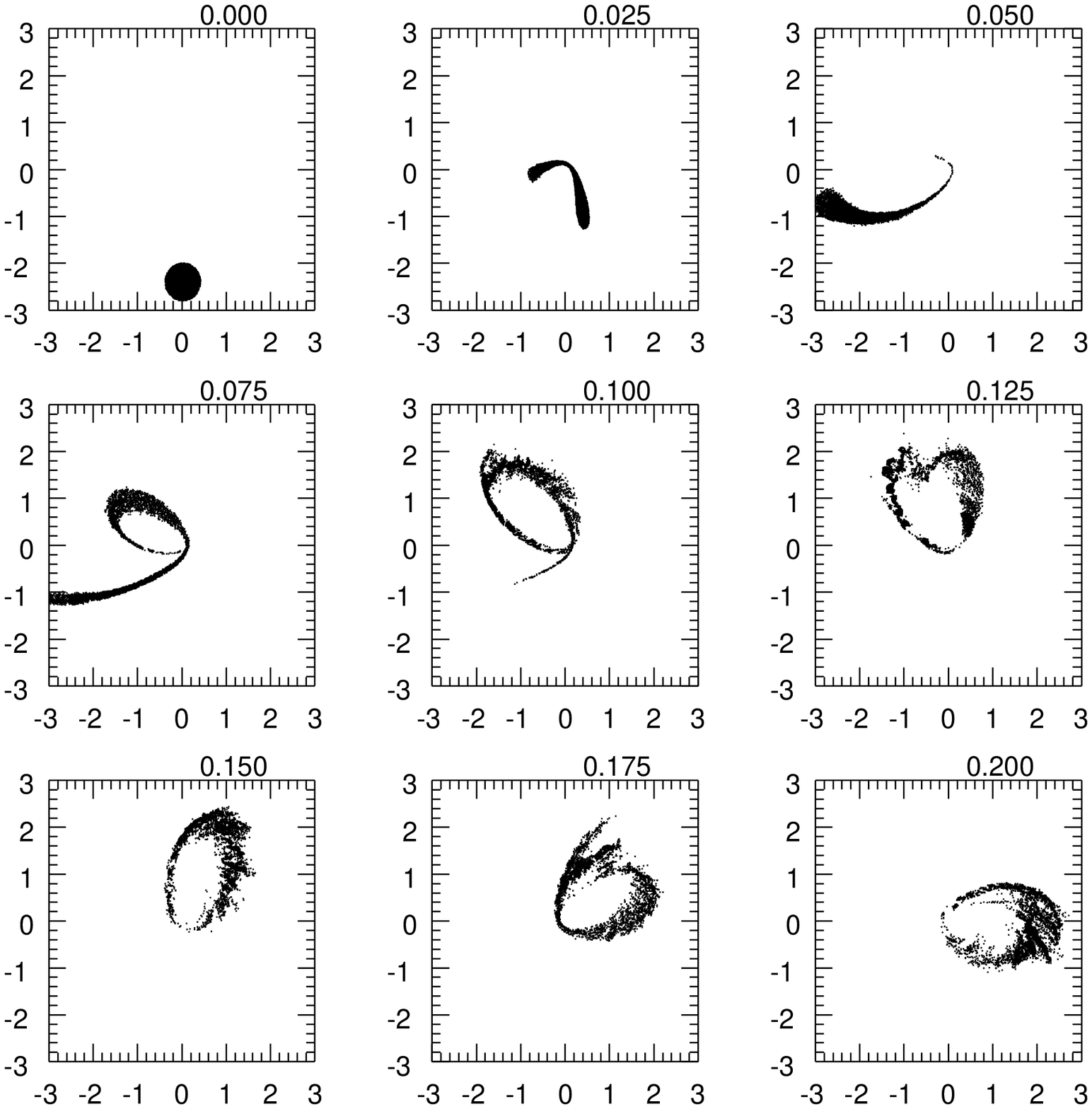}
\caption{Time evolution of a small cloud (from Paper 1), 
initially at a mean distance of
2.4 pc from the center, on a very low angular momentum orbit 
($V_t = 0.2V_c$).  The distance is in units of parsecs and 
the time
since beginning of the infall is given above each frame in units of
one-million years}
\end{figure}

The dispersion ring model maybe more relevant to the ionized filaments
within the central cavity of the CND.  In Fig.\ 3 we see the results of
the second calculation:  that 
of a small cloud, initially 2.0 to 2.8 pc from the center, on a very
low angular momentum orbit  ($V_t = 0.2 V_c$) which carries it within
0.2 pc of the center.  The cloud is initially tidally stretched into
a very long filament which repeatedly collides with itself.  By 
an epoch of $1.5\times 10^5$ years, a clear dispersion ring is formed.
This coherent and relatively long-lived structure 
precesses at an angular speed of -26 km s$^{-1}$pc$^{-1}$.  It is longer-lived
than the CND model in terms of orbit times because there is less 
differential precession due to the smaller initial size of the cloud.
In the eighth frame of Fig.\ 3 (t = $1.75\times 10^5$ years),
the cloud is well-described by the
dispersion orbits shown in Fig.\ 1d.  Note in particular the much higher
density of gas on the side where the material is approaching the center
due to the differential precession of streamlines, just as in Fig.\ 1d.  
This could explain why the Northern Arm is not observed to be a complete ring.

When this structure as it appears in the eighth frame is appropriately 
projected onto the plane of the sky (for this projection the orbital 
plane almost coincides
with that of the CND cloud), the morphology and distribution
of radial velocity again bears a strong resemblance to that observed
for the Northern Arm (Gezari et al. 1996, Herbst et al. 1993, Roberts
\& Goss 1993).  In Fig.\ 4 we see the morphology and the projection of
the velocity vectors of on the plane of the sky.  The sense 
of proper 
motion agrees with that recently observed by Yusef-Sadeh et al. (1998)
and by Zhao \& Goss (1998).
\begin{figure}
\plotone{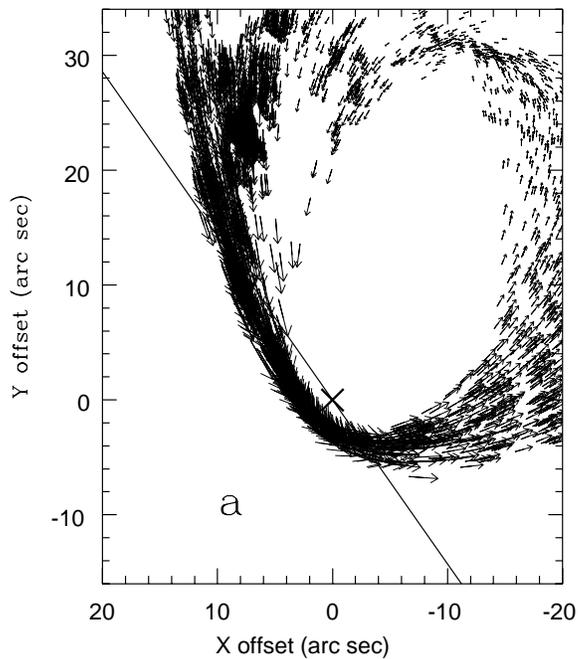}
\caption{The Northern Arm model as seen in the eighth frame of Fig.\ 3
($1.75\times 10^5$ years) projected onto the plane of the sky (from Paper 1).
This is a particle
plot where the arrows indicate the sense and magnitude of the proper motion. 
Coordinates are offsets in arc seconds from the dynamical center.
Here again the north-western side (upper right) is the near side; 
the plane of the Northern Arm model coincides with the plane of the 
CND model to within 10 degrees.  The model matches the observed morphology
and kinematics of the Northern Arm.  Note in particular the higher
density on the western side (due to orbit crowding as in Fig.\ 1d); 
this gives the appearance of an incomplete ring.}
\end{figure}

\section {Star formation in strong shocks}

While the dispersion ring is forming, the tidally stretched cloud collides
with itself near the point mass (Fig.\ 3, fourth and fifth frames). 
The collision velocities exceed 100 km/s.  Assuming 
highly efficient radiation of the thermal energy, which is likely in the
molecular gas, the resulting strong shocks will lead to
significant compression of the gas and high gas densities;
indeed the gas densities might well exceed the Roche limit 
($10^{13}$ to $10^{14}$ particles cm$^{-3}$) in the 
tidal field of a $2.5\times 10^6$ M$_\odot$ point mass at a distance
of 0.1 pc (Phinney 1989).
Such a mechanism might very well be the explanation for the 
young stars observed within a few tenths of a parsec of Sgr A$^*$, the
putative massive black hole at the Galactic Center (Allen et al. 1990,
Eckart et al. 1993, Krabbe et al. 1995).

The sticky particle algorithm applied here allows one to determine
the local value of the compression at the location of a particle
($-\nabla\cdot V$).  This may then be used as a criterion for star formation.
If the compression exceeds some arbitrary
threshold (in this case 2000 km s$^{-1}$pc$^{-1}$), that particular particle 
is re-tagged as a star;  thereafter, there are no ``viscous'' interactions
with other particles, and the particle's motion is determined only
by gravity.

The results of such a calculation are shown in Fig.\ 6 where we see both
the gas and star distribution after roughly one and two precession periods
of the dispersion ring.  We see that the gaseous dispersion ring
resembling the Northern Arm can easily persist for two precession periods
or roughly 10 to 20 characteristic orbital periods;  there is no need
for this structure to be a highly transient manifestation of a tidally
disrupting cloud on the first passage by the black hole.  Basically,
the dispersion ring appears as an ``attractor'' in the phase space
of the system-- a well-known phenomenon in dissipative systems.
The stars, on the other hand, without any such mechanism for 
self-organization, slowly diffuse throughout the available phase space.

\begin{figure}
\plotone{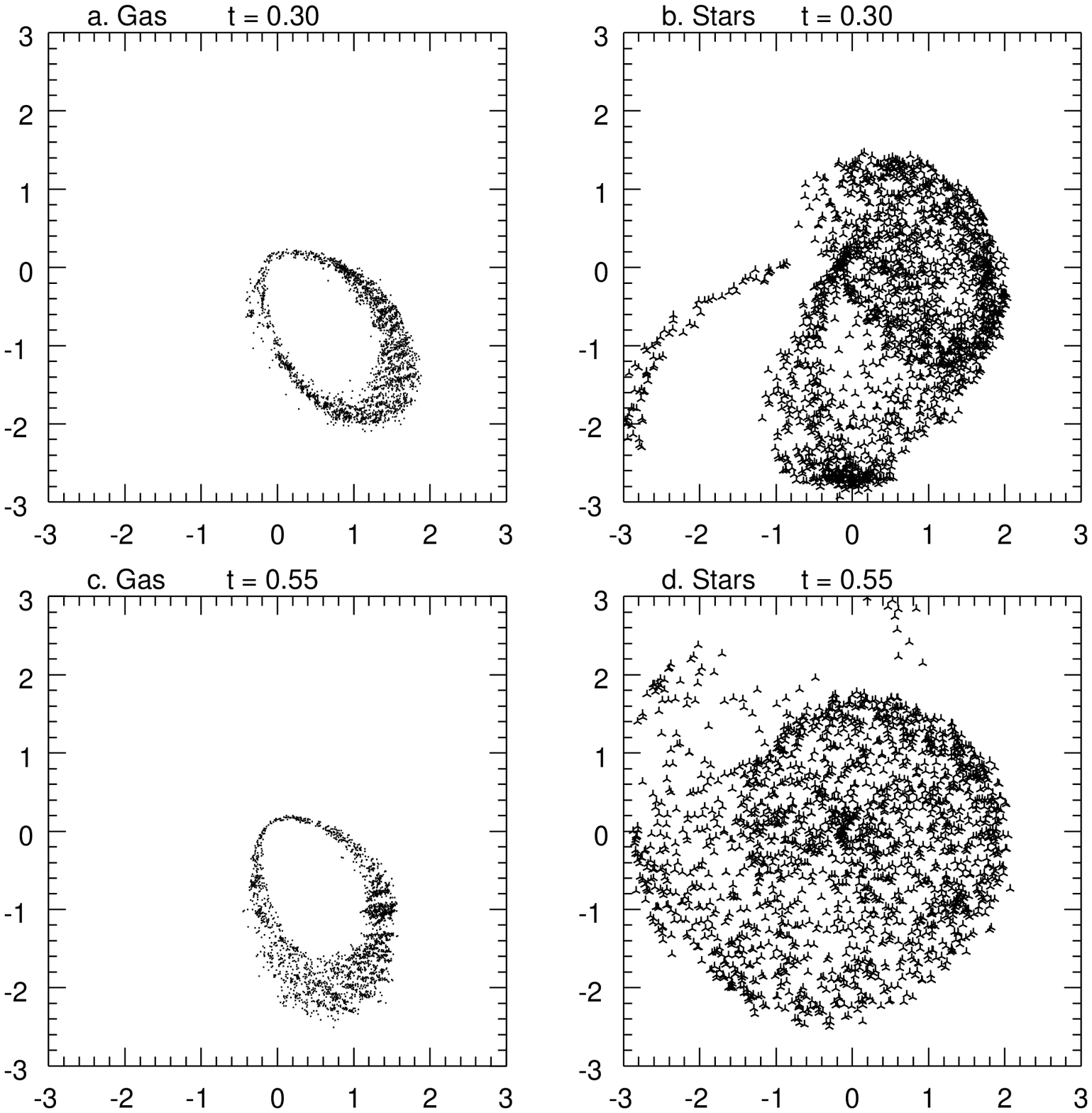}
\caption{a,b)  Gas and star distributions at t = $3\times 10^5$ years
corresponding to
more than one complete precession of the dispersion ring.  c,d) 
Gas and star distributions at t = $5.5 \times 10^5$ years 
or almost one precession
time later.  Distance is in units of pc.  This figure is from Paper 1.}
\end{figure}

\section{Comments}

We may conclude that the tidal disruption of a small cloud in a potential 
containing a point mass will lead to a long-lived gas structure--
a dispersion ring-- which precesses counter to the sense of 
gas rotation.  Such objects resemble, morphologically and kinematically, 
the gaseous structures seen in
the inner few parsecs of the Galaxy-- the CND and, within the central cavity
of the CND, the Northern Arm.  While forming a dispersion ring,
a low angular momentum cloud (passing near the central point mass),
is stretched into a long filament which collides with itself several times.
The resulting strong shocks lead to high compression and can be the 
sites of star-formation, even in the near tidal field of the black hole.
This may explain the presence of massive stars less than several million
years old within 0.1 pc of Sgr A$^*$.

Of course, I have not dealt with the problem of initial conditions.  How
are clouds launched on such orbits to begin with?  We know from a number
of observations (Morris \& Serabyn 1996) that the interstellar medium in
the inner 200 pc of the Galaxy is highly inhomogeneous and very turbulent--
most of the gas is found in massive molecular clouds and the random
velocities of these clouds are a considerable fraction of the circular
velocity.  The decay of supersonic turbulence occurs through cloud-cloud
collisions so occasionally, through this process, a low angular momentum
cloud will be created.    
Bursts of star formation or the occasional flaring of the black
hole are possible mechanisms for maintaining the turbulence.

Such a scenario might have general relevance to accretion onto massive 
black holes in active and normal galactic nuclei.  Accretion may
proceed via a series of such tidal disruptions of clouds on low angular
momentum orbits.  If this were true, we would expect accretion to be highly
episodic but also highly inefficient with most gas disappearing
into star formation.


\begin{references}

\reference Allen, D.A., Hyland, A.R., Hillier, D.J. 1990, MNRAS, 244, 706

\reference Eckart, A., Genzel, R., Hormann, R., Sams, B.J., Tacconi-Garman 1993,
  ApJ, L77

\reference G\"usten, R., Genzel, R., Wright, M.C.H., Jaffe, D.T., Stutzki, J.,
   Harris, A.I. 1987, ApJ, 318, 124

\reference Gezari, D., Dwek, E., Varosi, F. 1996, in IAU Symp.169, Unsolved
  Problems of the Milky Way (eds. L.\ Blitz \& P.\ Teuben),
  Dordrecht: Kluwer, 231

\reference Herbst, T.M., Beckwith, S.V.W., Forrest, W.J., Pipher, J.L. 1993,
  AJ, 105, 956

\reference Krabbe, A., Genzel, R., Eckart, A., Najarro, F., Lutz, D., 
  Cameron, M.,
  Kroker, H., Tacconi-Garman, L.E., Thatte, N., Weitzel, L., Drapatz, S.,
  Geballe, T., Sternberg, A., Kudritzki, R. 1995, ApJ, 447, L95

\reference Lindblad, B. 1956, Stockholms Obs.Ann., 19, No.\ 7

\reference Mezger, P.G., Duchl, W.J., Zylka, R. 1996, A\&A Rev.\, 7, 289

\reference Morris, M., Serabyn, E. 1996, Ann.Rev.A.Ap., 34, 645

\reference Phinney, E.S. 1988, in IAU Symp.136, The Center of the Galaxy
  (ed. M.\ Morris), Dordrecht: Kluwer, 543 

\reference Roberts, D.A., Goss, W.M. 1993, ApJS, 86, 113

\reference Roberts, D.A., Yusef-Zadeh, F., Goss, W.M. 1996, ApJ, 459, 627

\reference Sanders, R.H. 1998, MNRAS, 294, 35 (Paper 1)

\reference Syer, D., Clarke, C.J. 1992, MNRAS, 255, 92

\reference Serabyn, E., Lacy, J.H., Townes, C.H., Bharat, R. 1988, ApJ,
  326, 171

\reference Yusef-Zadeh, F., Roberts, D.A., Biretta, J. 1998, ApJ, 503, L191

\reference Zhao, J.-H., Goss, W.M. 1998, ApJ, 499, L163
\end{references}
\end{document}